\documentclass[sigconf, nonacm]{acmart}

\newcommand\vldbdoi{XX.XX/XXX.XX}
\newcommand\vldbpages{XXX-XXX}
\newcommand\vldbvolume{14}
\newcommand\vldbissue{1}
\newcommand\vldbyear{2025}
\newcommand\vldbauthors{\authors}
\newcommand\vldbtitle{\shorttitle} 
\newcommand\vldbavailabilityurl{https://github.com/gdymind/Buck-Learned-Index and https://github.com/wwl2755/GRE_buck_index}
\newcommand\vldbpagestyle{plain} 

\usepackage[ruled,vlined,linesnumbered]{algorithm2e}
\usepackage{amsmath}
\usepackage{float}
\usepackage{multirow}
 \usepackage{balance}
\SetKwComment{Comment}{/* }{ */}
\SetKwIF{If}{ElseIf}{Else}{if}{then}{else if}{else}{end if}
\SetKwFor{For}{for}{do}{end for}
\SetKwFor{While}{while}{do}{end while}
\SetKwFunction{ModelInsert}{ModelInsert}
\SetKwFunction{bottomUpPropagate}{bottomUpPropagate}

\begin{document}
\title{BLI: A High-performance Bucket-based Learned Index with Concurrency Support}

\author{Huibing Dong}
\affiliation{%
  \institution{University of Minnesota}
}
\email{dong0198@umn.edu}

\author{Wenlong Wang}
\affiliation{%
  \institution{University of Minnesota}
}
\email{wang9467@umn.edu}

\author{Chun Liu}
\orcid{0000-0001-5109-3700}
\affiliation{%
  \institution{ByteDance}
}
\email{chun.liu@bytedance.com}


\author{David Du}
\affiliation{%
  \institution{University of Minnesota}
}
\email{du@umn.edu}

\begin{abstract}
Learned indexes are promising to replace traditional tree-based indexes. They typically employ machine learning models to efficiently predict target positions in strictly sorted linear arrays. However, the strict sorted order 1) significantly increases insertion overhead, 2) makes it challenging to support lock-free concurrency, and 3) harms in-node lookup/insertion efficiency due to model inaccuracy.\

In this paper, we introduce a \textit{Bucket-based Learned Index (BLI)}, which is an updatable in-memory learned index that adopts a "globally sorted, locally unsorted" approach by replacing linear sorted arrays with \textit{Buckets}. BLI optimizes the insertion throughput by only sorting Buckets, not the key-value pairs within a Bucket. BLI strategically balances three critical performance metrics: tree fanouts, lookup/insert latency for inner nodes, lookup/insert latency for leaf nodes, and memory consumption. To minimize maintenance costs, BLI performs lightweight bulk loading, insert, node scaling, node split, model retraining, and node merging adaptively. BLI supports lock-free concurrency thanks to the unsorted design with Buckets. Our results show that BLI achieves up to 2.21x better throughput than state-of-the-art learned indexes, with up to 3.91x gains under multi-threaded conditions.
\end{abstract}

\maketitle

\pagestyle{\vldbpagestyle}
\begingroup\small\noindent\raggedright\textbf{PVLDB Reference Format:}\\
\vldbauthors. \vldbtitle. PVLDB, \vldbvolume(\vldbissue): \vldbpages, \vldbyear.\\
\href{https://doi.org/\vldbdoi}{doi:\vldbdoi}
\endgroup
\begingroup
\renewcommand\thefootnote{}\footnote{\noindent
This work is licensed under the Creative Commons BY-NC-ND 4.0 International License. Visit \url{https://creativecommons.org/licenses/by-nc-nd/4.0/} to view a copy of this license. For any use beyond those covered by this license, obtain permission by emailing \href{mailto:info@vldb.org}{info@vldb.org}. Copyright is held by the owner/author(s). Publication rights licensed to the VLDB Endowment. \\
\raggedright Proceedings of the VLDB Endowment, Vol. \vldbvolume, No. \vldbissue\ %
ISSN 2150-8097. \\
\href{https://doi.org/\vldbdoi}{doi:\vldbdoi} \\
}\addtocounter{footnote}{-1}\endgroup

\ifdefempty{\vldbavailabilityurl}{}{
\vspace{.3cm}
\begingroup\small\noindent\raggedright\textbf{PVLDB Artifact Availability:}\\
The source code, data, and/or other artifacts have been made available at \url{\vldbavailabilityurl}.
\endgroup
}

\section{Introduction \label{section-intro}}

In-memory key-value stores are widely used across various data-centric applications, as evidenced by the key-value store deployment at Meta\cite{nishtala2013scaling}, Twitter\cite{yang2021large}, and other companies. While tree-based indexes, such as B+-trees, have traditionally dominated the indexing of key-value stores which support range queries, recent advancements have introduced learned indexes (e.g., RMI\cite{kraska2018case}, ALEX\cite{ding2020alex} and LIPP\cite{wu2021updatable}) as superior alternatives due to their enhanced or comparable lookup performance\cite{wu2021updatable, kraska2018case, wang2024learnedkv}. Learned indexes use machine learning models to predict the approximate position of a lookup key in a strictly sorted array of key-value pairs. Trained on the mapping from keys to their positions in the array, these models aim for fast and accurate prediction, outperforming traditional tree-based index searches. For simplicity, we refer to the distribution of all (key, position) pairs as the \textit{key distribution} hereafter. If the exact key-value pair is not found at the predicted position, a last-mile search is performed in nearby positions of the linear array. The \textit{key-value pairs are strictly sorted in the array}, and the learned models act as monotonic functions of the key-position mapping.

The first proposed learned index, the Recursive Model Index (RMI)\cite{kraska2018case}, employs a hierarchy of models for enhanced accuracy. The root model spans the entire key range, branching into child models, each dedicated to a sub-key range. This branching process concludes after multiple iterations, forming multiple levels of models. After that, the last-level models (or the leaf models) predict the location of a specific key-value pair within its key range.

However, key-value pairs in RMI are tightly packed into a dense array, which precludes accommodating new insertions within the array. Therefore, various \textit{updatable learned indexes} have been developed to support insertions of key-value pairs, and the number of models can be dynamically increased accordingly. Thanks to the "sufficiently accurate" model predictions, the throughput of existing updatable learned indexes can outperform traditional indexes by over 80\%, according to a benchmarking paper called GRE\cite{wongkham2022updatable}. These updatable learned indexes typically utilize one of three structures to facilitate insertions: merge trees, delta buffers, or gapped arrays. During insertion, the merge tree method merges multiple read-only sub-indexes with newly inserted key-value pairs. In the delta-buffer approach, new insertions are temporarily held in a buffer area, which will be periodically merged into the main index. The gapped-array strategy employs pre-allocated spaces (i.e., empty positions in the linear array) to accommodate new key-value pairs in place. We refer to gapped-array-based learned indexes as \textit{in-place updatable learned indexes}. 
Notably, according to GRE, two state-of-the-art (SOTA) in-place updatable learned indexes, ALEX\cite{ding2020alex} and LIPP\cite{wu2021updatable}, generally outperform other learned indexes. ALEX supports insertions by shifting neighboring key-value pairs within a leaf model's key range. 
GRE's analysis reveals that, during insertion, ALEX spends an average of 58.60\% of its time on shifting operations after the lookup process, while LIPP spends 32.24\% on chaining operations. These findings motivate us to further improve performance by eliminating strict-order-preserving operations such as shifting and chaining. We introduce a new in-place updatable learned index, the \textit{Bucket-Based Learned Index (BLI)}, which operates without strict ordering. Before diving into the specifics of BLI, we will introduce some key performance metrics to quantify potential optimizations.

\subsection{Performance metrics}
As aforementioned, learned indexes generally use a hierarchy of machine learning models. A non-leaf model holds an array of child models, while a leaf model holds an array of key-value pairs within its key range. The overall lookup and insertion performance is influenced by both the efficiency of individual models and the number of models that need to be accessed during these operations. We have defined four key metrics that impact the overall performance:

\textbf{1) Model Fanout ($N$)}. The Model Fanout refers to the number of child models in a non-leaf model or the number of key-value pairs in a leaf model. Given the same dataset, the lookup traversal path (i.e., the sequence of all models from the root to the leaf) can be shortened if larger model fanouts are accommodated.

\textbf{2) Lookup Efficiency ($E_{\text{lookup}}$).}
For an individual model, given $t_{\text{lookup}_i}$ representing the lookup latency when searching for the $i$th child model or key-value pair in the model's array, lookup efficiency $E_{\text{lookup}}$ can be defined as the reciprocal of the average lookup latency of the entries (i.e., key-value pairs or child models) corresponding to the model, expressed as $E_{\text{lookup}} = \frac{N}{\sum_{i=1}^N(t_{\text{lookup}_i})}$. $E_{\text{lookup}}$ is determined by all lookup latencies $t_{\text{lookup}_i}$ with $i$ from $1$ to $N$, and these latencies are mainly determined by the model prediction error. Balancing the tradeoff between minimizing prediction error in $E_{\text{lookup}}$ and maintaining an optimal $N$ is crucial. While a low prediction error improves $E_{\text{lookup}}$, excessively reducing the error tolerance can decrease $N$, which is detrimental.

\textbf{3) Insert Efficiency ($E_{\text{insert}}$).} 
Before performing insertions, a lookup is required to determine the appropriate insertion position, after which insertions can be executed correctly.  To exclude the effect of lookup efficiency, we define $E_{\text{insert}}$ as $\frac{N}{\sum_{i=1}^N(t_{\text{insert}_i} - t_{\text{lookup}_i})}$, where $t_{\text{insert}_i}$ represents the total latency during insertion.

\textbf{4) Memory Overhead ($O_{\text{mem}}$).} Memory overhead is defined as the ratio of the memory usage of the entire system (including both key-value pairs and the index) to the size of all key-value pairs. For example, if all data is 10GB and the memory usage of the system is 15GB, then the memory overhead is $O_{\text{mem}} = \frac{15}{10} = 150\%$.

Existing in-place updatable learned indexes often suffer from degraded $E_{\text{insert}}$ due to strict-order-preserving operations such as shifting in ALEX and chaining in LIPP. However, we find that by relaxing the strict order constraint, the four metrics (i.e., $N$, $E_{\text{lookup}}$, $E_{\text{insert}}$, and $O_{\text{mem}}$) can be further balanced and optimized, leading to improved lookup and insertion performance.

\subsection{Our proposed Bucket design \label{sec-intro-design}}
We present BLI, which employs a "globally sorted, locally unsorted" design to address several key issues in existing learned indexes. This approach utilizes a hierarchical structure of machine learning models, which is a common architecture in most existing learned indexes. Unlike these indexes, BLI does not require the sorting of key-value pairs within the key range of a leaf model. Instead, each leaf model in BLI groups unsorted key-value pairs into a \textit{ Bucket}. However, Buckets are sorted based on their non-overlapping key ranges, thus, embodying the "globally sorted, locally unsorted" concept. 
We summarize our contributions as follows:

\begin{itemize}
    \item \textbf{Improved insertion efficiency with "locally unsorted" design.} Key-value pairs are unsorted within a Bucket. A new key-value pair will be inserted into its Bucket without strict-order-preserving operations such as shifting and chaining, thus significantly enhancing insertion efficiency. In addition, BLI provides a suggested location within a Bucket for each key-value pair insertion to maintain a comparable lookup efficiency with state-of-the-ar learned indexes.
    \item \textbf{Simplified key distribution with larger prediction units.} Unlike existing indexes, each Bucket in BLI has a representative key chosen from its corresponding key range, and prediction models to indicate the target Bucket based on its representative key rather than a slot holding an individual key-value pair.  As a result, the key distribution can be more linear when compared to predicting the individual keys, as will be proved in \autoref{sec-model-simplification}, which can lead to reduced prediction errors.
    \item \textbf{Consistent model accuracy over its lifetime.} Existing learned indexes typically update their models periodically to avoid high retraining overhead. However, as more insertions occur over time, the model's accuracy gradually decreases, which can negatively impact $E_{\text{lookup}}$. BLI's models predict the target Bucket rather than individual keys, reducing the frequency of model updates when inserting key-value pairs into Buckets and thus maintaining a better consistent model accuracy.
    \item \textbf{Tunable perfomance with N, $E_\text{lookup}$, $E_\text{insert}$, and $O_{\text{mem}}$.} We identified four major metrics, model fanout $N$, lookup efficiency $E_\text{lookup}$, insert efficiency $E_\text{insert}$, and memory overhead $O_{\text{mem}}$, that affect the performance of learned indexes. Moreover, we investigated BLI's ability to make tradeoffs among these metrics, making BLI adaptive to different workloads.
    \item \textbf{Linear-disbtribution-aware neighbor merging: } With new key-value pairs inserted, the key distribution of adjacent models may become "linear enough" at some point, and BLI is able to merge those adjacent models into a single linear model.
    \item \textbf{Lock-free concurrency support:} Supporting lock-free concurrency in in-place updatable learned indexes is challenging since operations to enforce strictly sorted order (such as shifting and chaining) require locking to guarantee data consistency. BLI supports lock-free concurrency with the single-producer-multi-consumer threading assumption. We employ a bitmap to indicate valid key-value pairs and read-copy-update to ensure correctness.
\end{itemize}

As a result, our proposed BLI surpasses the throughput of state-of-the-art learned indexes by 1.94 times. In multi-threaded scenarios, BLI outperforms other systems by up to 3.91 times.

The structure of this paper is as follows: Section \ref{sec-realted-work} reviews related work and provides a brief comparison with BLI. Section \ref{sec-motivation} identifies key research challenges posed by the strictly sorted order in existing learned indexes, supported by motivational experiments and analysis. Section \ref{sec-architecture} introduces the proposed BLI architecture based on the "globally sorted, locally unsorted" design. Section \ref{sec-algorithm} details the main operations of BLI, including lookup, insertion, bulk loading, and concurrency support. Section \ref{sec-evalution} presents a comparative evaluation of BLI against existing learned indexes and offers an in-depth analysis of BLI's performance. Finally, Section \ref{sec-conclusion} summarizes the overall contributions of this work.

\section{Related work \label{sec-realted-work}}
This section will first outline the evolution from read-only to updatable learned indexes, then discuss the main operations in updatable learned indexes, and briefly compare existing methods with BLI.

\subsection{Evolution from read-only to updatable learned indexes}
The first learned index, Recursive Model Index (RMI)\cite{kraska2018case}, utilizes a hierarchy of models to improve prediction accuracy. RMI starts with the root learned model that covers the entire key range and then branches the root into child models that each handle a subset of the key range. This hierarchical division continues through several iterations, ultimately creating multiple levels of models with good prediction accuracy. At the last level, a leaf model makes predictions only on the keys within its corresponding sub-key range.

However, RMI was read-only since it was built atop a static set of data and a densely packed array. Later, \textit{updatable learned indexes} were introduced to accommodate new key-value pair insertions. In updatable learned indexes, the global sorted array can further split into multiple sub-arrays corresponding to the key ranges of their leaf models. These sub-arrays of key-value pairs are stored along with their leaf models for ease of maintenance. Existing approaches typically utilize the following three techniques \cite{wongkham2022updatable} to accommodate insertions. 1) \textbf{Merge Trees}: PGM-Index\cite{ferragina2020pgm} employs a collection of static sub-learned indexes. During insertions, the newly inserted key-value pair, along with a subset of sub-learned indexes, are merged in a write-optimized LSM-tree style, which compromises the lookup performance since all sub-learned indexes need to be visited. 2) \textbf{Delta Buffers}: XIndex\cite{tang2020xindex} and FINEdex\cite{li2021finedex} place new insertions temporarily to some separate area called delta buffers. However, these designs also degrade lookup speeds as it is necessary to check across both the major learned index and the delta buffer. 3) \textbf{Gapped Arrays}: ALEX\cite{ding2020alex} and LIPP\cite{wu2021updatable} incorporate pre-allocated gaps (i.e., empty slots) to integrate incoming key-value pairs. Typically, these in-place approaches perform better since the merge-tree and delta-buffer-based methods involve accessing multiple structures (sub-indexes or delta buffers) during a lookup. According to GRE, two state-of-the-art gapped-array-based learned indexes, ALEX \cite{ding2020alex} and LIPP \cite{wu2021updatable}, generally outperform other existing updatable learned indexes, and we mainly focus our comparisons with these two systems. Subsequent sub-sections will detail the main operations of existing updatable learned indexes, especially ALEX and LIPP, and discuss how our BLI design can offer improvements.

\subsection{Lookup: predicting an individual key vs. a key range}
The lookup process can typically be split into a \textit{traverse-to-leaf lookup}, which selects the appropriate sub-key range and traverses to the corresponding leaf model, and a \textit{leaf lookup}, which uses the leaf model to predict the position of key-value pairs. Existing learned indexes \cite{ferragina2020pgm, tang2020xindex, li2021finedex, galakatos2019fiting, ding2020alex, wongkham2022updatable} predict the offset of a target key within the corresponding array of the model.

Unlike most existing updatable learned indexes, our proposed BLI focuses on a different dimension: prediction granularity.
BLI predicts a key range, which is represented by a Bucket rather than an individual key. The Bucket layout allows for efficient last-mile search by bypassing all entries from non-target Buckets. Furthermore, the prediction model is more linear since Buckets simplifies the key distribution, which will be proved in Section \ref{sec-resilience-model-inaccuracy}.

\subsection{Insertion: strictly sorted order vs. "globally sorted, locally unsorted" order }
We focus on in-place updatable learned indexes, especially ALEX and LIPP, since they generally outperform other updatable learned indexes. Specifically, ALEX reserves empty slots in arrays corresponding to its leaf models to facilitate insertions. When a new key-value pair is inserted using a leaf model, if the model-predicted slot is occupied, ALEX shifts neighboring pairs to nearby gaps to maintain a strict sorted order. 
Conversely, LIPP does not distinguish between non-leaf and leaf models, allowing the placement of key-value pairs, child pointers, and empty slots throughout any model in its hierarchy. During insertion, if a predicted slot is occupied, LIPP resorts to chaining, 
by creating a new child model containing both the existing and new pairs, subsequently replacing the existing pair with a pointer to this new child model.

To ensure high lookup and insertion performance, our proposed learned index, BLI, also supports in-place insertions through pre-allocated gaps in arrays corresponding to their leaf models. Unlike ALEX, which requires data insertion into strictly sorted linear arrays, BLI incorporates empty slots within each Bucket. Since keys within a Bucket are unsorted, BLI avoids the overhead associated with shifting operations. Moreover, while LIPP requires chaining for every occupied predicted position, BLI simply searches forward to find the next available empty slot, benefiting from the unsorted nature of Buckets.

Notably, insertions in all updatable learned indexes necessitate Structure-Modification Operations (SMOs), such as model splitting, scaling, or retraining, triggered by capacity limits or performance degradation. Efficient execution and the trigger conditions for these SMOs are crucial for maintaining system performance. Moreover, the way SMOs are implemented also affects the three aforementioned metrics affecting performance (i.e., the model fanout $N$, in-model lookup efficiency $E_{\text{lookup}}$, and in-model insert efficiency($E_{\text{insert}}$).\

\subsection{Bulk loading: top-down vs. bottom-up}
Bulk loading is a process that builds the whole learned index given a sorted array of key-value pairs. 
Bulk loading can be implemented using a top-down or bottom-up method. Given a sorted linear array of key-value pairs, top-down methods\cite{ding2020alex, wu2021updatable} start by generating the root model and then recursively branching into child models. In contrast, bottom-up bulk loading methods \cite{ferragina2020pgm, tang2020xindex, li2021finedex} first partition all key-value pairs into disjoint key ranges and train leaf models for each key range. After that, each key range is typically represented by its minimum key and a pointer to this key range. The (\texttt{min key}, \texttt{pointer}) pairs propagate upwards in the same way until a single model (i.e., the root) is generated.

Both ALEX and LIPP adopt a top-down method. To train the root model, they transverse every key-value pair in the sorted array. The traversing of the training process happens in each newly-generate sub-key range. As a result, they traverse the data multiple passes, which can typically be avoided in a bottom-up approach.

BLI employs a bottom-up approach to construct the tree. In a single pass, BLI segments key-value pairs into Buckets and then operates on the (\texttt{min key}, \texttt{pointer}) iteratively. Similar to the existing bottom-up approaches, although the upper-level models also need to transverse the (\texttt{min key}, \texttt{pointer}) pairs for training, the number of such pairs decreases exponentially after each iteration. As a result, only a single pass of data traversal is required in BLI. After bulk loading, the BLI structure has no long or skewed paths.

\subsection{Concurrency: use locking to ensure conrrectness}
Previous work has relied on locking mechanisms to support concurrency. XIndex \cite{tang2020xindex} and FINEdex \cite{li2021finedex} use temporary buffers to accommodate new key-value pairs, periodically merging the buffered data into the main index structure. XIndex applies locks to individual records within the data array and uses a per-group lock for sequential insertions. FINEdex avoids locks during reads by employing version control, but it still requires locks during insertions. SALI \cite{ge2023sali} employs per-node write locks on LIPP. The locking mechanism blocks user requests in the event of read or write conflicts. In contrast, thanks to the unsortedness of Buckets, BLI is able to support lock-free concurrency by utilizing valid bits and Read-Copy-Updates.

\section{Motivational Experiments and Analyses\label{sec-motivation}}
This section highlights some research issues associated with the strictly sorted order in existing learned indexes through motivational experiments and analyses. We examine their major operations (including lookup and insertion) and concurrency support.

\subsection{Lookup: prediction and local search efficiency \label{sec-resilience-model-inaccuracy}}
As previously mentioned, during a lookup operation, a model prediction in learned indexes is typically followed by a last-mile search to correct any prediction errors. Therefore, the aforementioned in-model lookup efficiency($E_{\text{lookup}}$) can be expanded as $E_{\text{lookup}} = \frac{N}{\sum_{i=1}^N(t_{\text{lookup}_i})} = \frac{N}{\sum_{i=1}^N(t_{\text{predict}_i} + t_{\text{local search}_i})}$. It is important to optimize and balance both $t_{\text{predict}}$ and $t_{\text{local search}}$ in order to maximize $E_{\text{lookup}}$.

\subsubsection{Model prediction \label{sec-model-simplification}}
Existing learned indexes utilize models to predict the position of an \textit{individual} \textit{key}, whereas our proposed BLI predicts a Bucket corresponding to a key-range. In other words, existing learned models typically employ \textit{key prediction models}, while our method utilizes \textit{range prediction models}. We argue that range prediction models result in equal or smaller errors compared to key prediction models. To elaborate, consider a sorted array of $N$ key-value pairs, where a key prediction model $m(k) = ak + b$ is trained, with $k$ as the input key, $a$ as the slope, and $b$ as the intercept. The parameters $a$ and $b$ are learned from the (\texttt{key}, \texttt{position}) distribution. For a (\texttt{key}, \texttt{position}) pair $(k, p)$, the prediction error for the key prediction model can be denoted as $e = \left|m(k) - p\right|$. Following the "globally sorted, locally unsorted" design, we can partition the $N$ key-value pairs into several groups, where each group contains $n$ adjacent key-value pairs, and the key ranges of these groups are not overlapped. We then define the range prediction model as $M(k)=\frac{m(k)}{n}$. That is, the training process for the range prediction model $M(k)$ is the same as that for the key prediction model $m(k)$, while the output of $M(k)$ is divided by the group size. Consequently, the prediction error for $M(k)$ becomes $E = \frac{\left|m(k) - p\right|}{n}$. Therefore, $E = \frac{e}{n} \leq e$, indicating that predicting a key range results in equal or smaller errors compared to predicting the position of an individual key. The gap between key range prediction error and individual prediction error is influenced by the group size or Bucket size in our context. A larger group or Bucket size can lead to reduced prediction errors. However, the Bucket size also affects the efficiency of the subsequent local search step, presenting a tradeoff between model prediction efficiency and local search efficiency. 

\begin{figure}[htbp]
\centering
\includegraphics[width=0.45\textwidth]{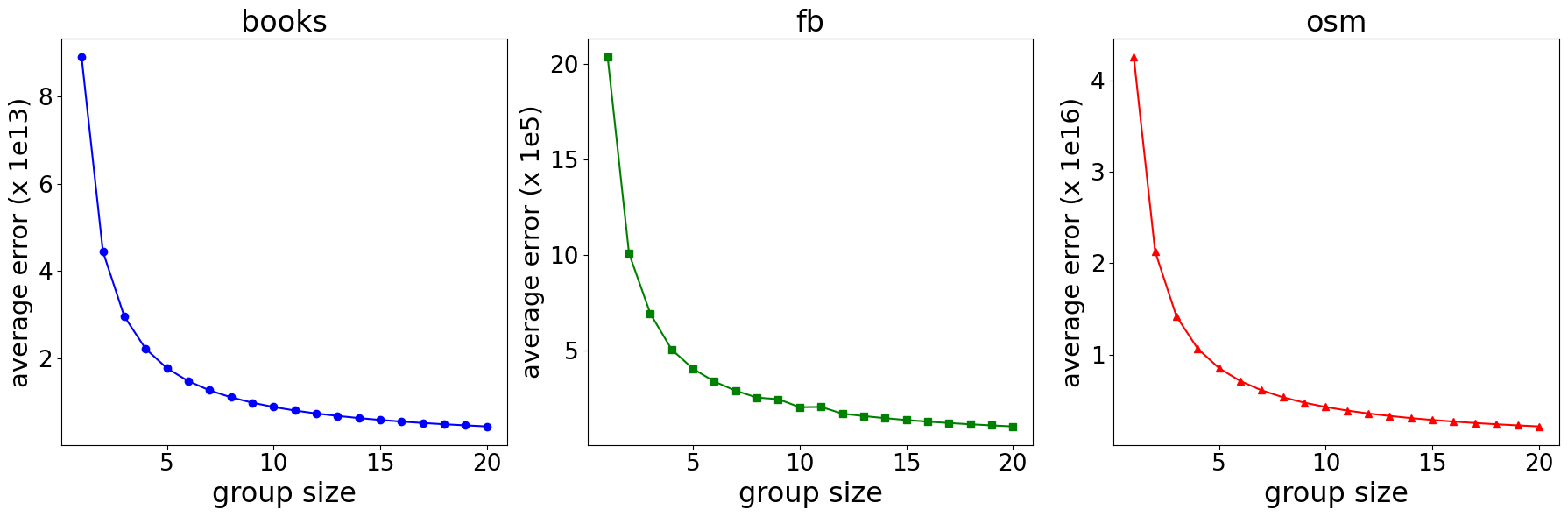}
\caption{\label{fig-err-vs-grp-size} The average prediction error with different group sizes on the first 10000 keys of \texttt{books}, \texttt{fb}, and \texttt{osm}.}
\end{figure}

To illustrate the impact of group or Bucket size on prediction errors, we conducted an experiment to show how prediction errors can be reduced. We trained a linear regression model on the first 10000 keys of the datasets \texttt{books}, \texttt{fb}, and \texttt{osm} with various group sizes. As shown in Figure \ref{fig-err-vs-grp-size}, we observed a significant reduction in the average prediction error as the group size increased.

\subsubsection{Last-mile search}
Existing learned indexes typically conduct local searches with the following three methods: 1) binary search with error bound\cite{kraska2018case, ferragina2020pgm, li2021finedex}, 2) exponential search with model-based insertion\cite{ding2020alex}, and 3) enforced precise predictions with chaining \cite{wu2021updatable}. Specifically,  some learned indexes using the first approach enforce the maximum prediction error for each learned model and will employ more learned models if the current ones cannot ensure the bounded error. The error bound is a predefined parameter of the system. After a model prediction, these learned index systems will perform a binary search on the neighboring key-value pairs within the error bound. ALEX, using the second method, introduces model-based insertion. It tries to insert a key-value pair to the model-predicted position if it has not been occupied. During lookup, if the predicted position contains a non-target key-value pair, it conducts an exponential search starting from the predicted location without bounded error, which could terminate at the end of the sorted array. LIPP, using the third method, also tries to insert the key-value pairs into the model-predicted positions, and it enforces precise prediction by conducting chaining operations of prediction models whenever there are prediction conflicts. As a result, LIPP eliminates the need to search its neighboring positions. On the other hand, it could result in a long chain of prediction models. Unlike existing approaches, our proposed BLI bounds its local searches within each Bucket.

\subsection{Insertion: overhead due to strict-order-preserving operations \label{sec-perfect-insert-overhead}}
Existing updatable learned indexes typically utilize strict-order-preserving operations to keep the key-value pair array sorted
We validate the negative effect of the extra operations by analyzing the time breakdown of ALEX and LIPP under an insert-only workload 
on three popular benchmarking datasets: \texttt{books}, \texttt{fb}, and \texttt{osm}. As observed, after the lookup process, ALEX spends 58.60\% of its time on shifting neighbors on average, while LIPP spends 32.24\% on average for chaining operations. These operations could be eliminated in our proposed "glocally sorted, locally unsorted" bucket design. On the other hand, the bucket design brings new tradeoffs that must be carefully handled.

\subsection{Concurrency}
Among learned indexes, FINEdex, XIndex, and PGM-Index support concurrency but require the use of delta buffers or merge trees. This reliance on multiple structures, such as delta buffers or sub-learned indexes, reduces lookup performance and introduces challenges in read-write workloads. Other state-of-the-art learned indexes, including ALEX and LIPP, need costly locking mechanisms to ensure data consistency in their strict-order layouts when enabling concurrency. For example, GRE implemented multi-threaded versions of ALEX and LIPP, known as ALEX+ and LIPP+, respectively, using locks. To evaluate the lock overhead, we compared the performance of ALEX with ALEX+ and LIPP with LIPP+. We began by bulk-loading 100 million key-value pairs and then executed 50 million random reads and 50 million random writes to measure throughput. Under a \textit{single} thread environment with the dataset \texttt{books}, ALEX+ exhibited a 20.10\% performance degradation, while LIPP+ showed a 26.75\% degradation compared to their original versions.

\section{BLI architecture \label{sec-architecture}}
This section presents the proposed BLI architecture derived from the "globally sorted, locally unsorted" design. BLI consists of sorted linear arrays of Buckets that contain unsorted data. These Buckets enhance insert efficiency while introducing new tradeoffs.

\subsection{Architecture overview}

Existing learned indexes typically use a hierarchy of machine-learning models to make "good enough" predictions. The lookup process starts at the root model, which covers the entire key range, and each model predicts the position of the next child model, narrowing the key range progressively.

\begin{figure}[htbp]
\centering
\includegraphics[width=0.48\textwidth]{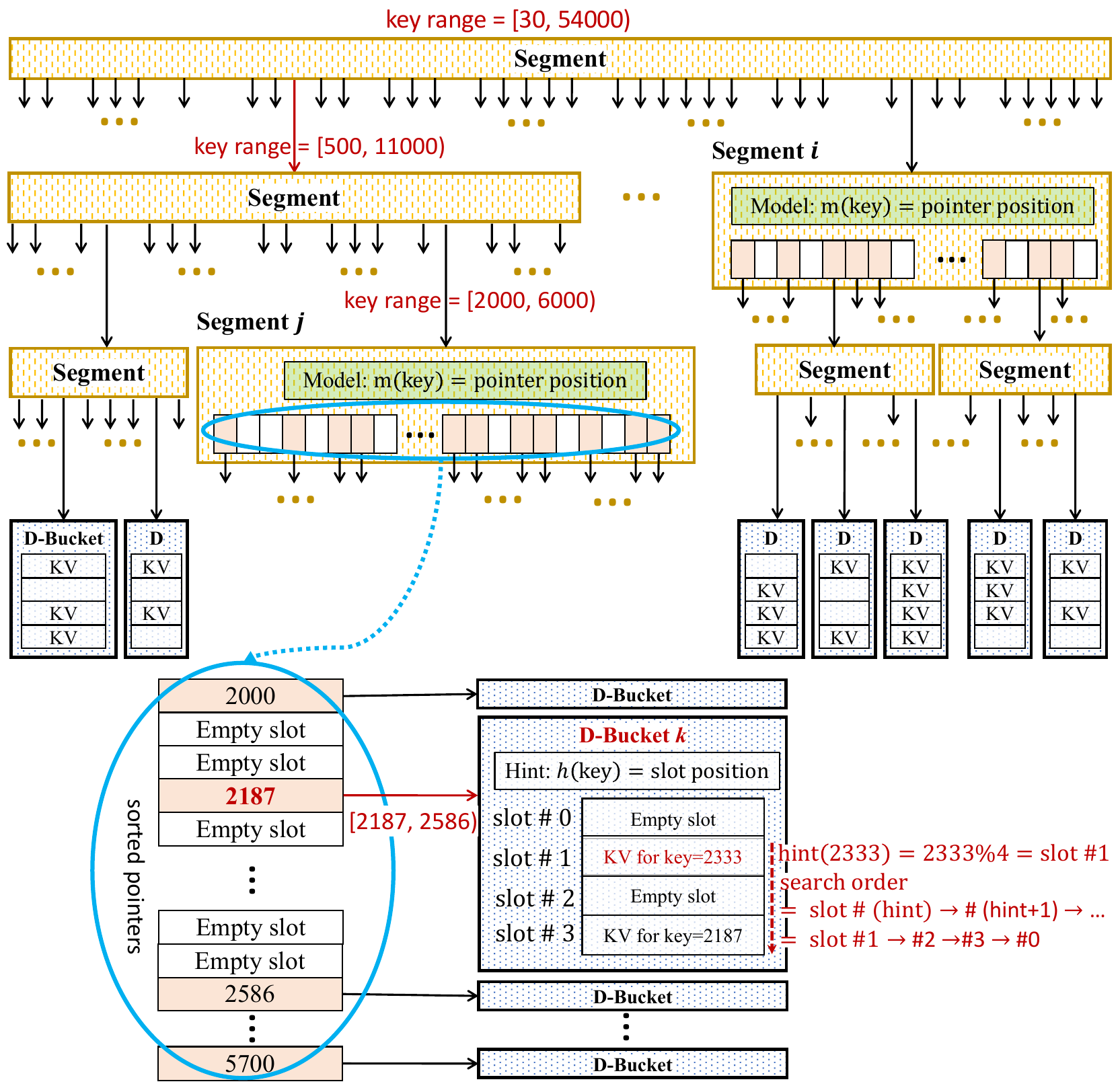}
\caption{\label{fig-architecture-base} An example of the BLI architecture, illustrating a three-level Segment and the bottom-level D-Bucket structure, employing the MOD hash as a hint function. }
\end{figure}

Adopting this hierarchical design, our proposed BLI features a sorted linear array of Buckets containing unsorted data. Figure \ref{fig-architecture-base} illustrates an example of the BLI architecture with three levels of \textit{Segments}, with the potential for more levels if additional data are present. A Segment spans a specific key range and includes a learned prediction model along with a sorted linear array of pointers to its child Segments, each responsible for a non-overlapping sub-key range. The learned model predicts the position of the child Segment based on a target key, and BLI searches nearby if the prediction has an error. At the last level, BLI organizes all key-value pairs within each sub-key range into a group called a \textit{Data Bucket} or \textit{D-Bucket}, where the key-value pairs are unsorted, but their positions will be suggested by a hint function. Empty slots can exist in segments and D-Buckets to facilitate future insertions.

Internal and last-level Segments share similar structures. The key difference is that internal Segments, such as Segment $i$ in Figure \ref{fig-architecture-base}, point to child Segments, whereas last-level Segments, like Segment $j$, point to D-Buckets. For instance, Segment $j$, responsible for the key range $[2000, 6000)$, comprises a model $m(\text{key})$ and sorted pointers covering its sub-key ranges: $[2000, 2187)$, $[2187, 2586)$, $\cdots$, $[5700, 6000)$. Specifically, D-Bucket $k$, handling the sub-key range $[2187, 2586)$, contains two key-value pairs and two empty slots.  The key-value pairs with keys $2187$ and $2333$ are placed in their respective hinted slots $1$ and $3$ using a MOD hash $h(\text{key}) = \text{key} \% 4$.

When looking up the key $2333$, BLI starts at the root Segment, which covers the entire key range $[30, 54000)$. The root Segment uses its prediction model to identify the target child Segment that spans the key range $[500, 11000)$, and then $[2000, 6000)$. This process is repeated in Segment $j$, and BLI ultimately reaches D-Bucket $k$. Once there, BLI applies the hint function $\text{hint}(2333) = \text{slot} 1$ as the starting search position. It continues searching through slots $2$, $3$, and $0$ until the target key-value pair is found if slot1 does not hold the key or all slots have been checked. In this example, slot $1$ is the correct position, so no further searching is required.

\subsection{Analysis of D-Buckets}
There are three design factors for D-Buckets: the size of D-Buckets, their fill ratio, and the hint design. These factors will influence the key aforementioned metrics: model fanouts ($N$), in-model lookup efficiency ($E_{\text{lookup}}$), in-model insert efficiency ($E_{\text{insert}}$), and memory overhead ($O_{\text{mem}}$). To fully optimize BLI, these factors should be balanced according to workload patterns (e.g., read-write ratio) and the system's memory capacity.

\subsubsection{D-Bucket size and the initial fill ratio}
\textit{D-Bucket size} refers to the number of slots within a D-Bucket, while the \textit{fill ratio} represents the ratio between the number of existing key-value pairs and the D-Bucket size. For example, D-Bucket $k$ in Figure \ref{fig-architecture-base} has four slots holding two key-value pairs, giving it a D-Bucket size of 4 and a fill ratio of 50\%. We pre-define the \textit{initial fill ratio} as the fill ratio when creating a new D-Bucket. The number of key-value pairs BLI will put in a newly created D-Bucket is equal to the Bucket size times the pre-defined initial fill ratio. For simplicity, the current BLI design fixes the D-Bucket size across the entire system. However, using variable D-Bucket sizes may lead to a more linear key distribution. The D-Bucket size influences the system in the following ways:
\begin{itemize}
\item With a larger D-Bucket, more key-value pairs can be placed into a single D-Bucket, which with the same model fanouts ($N$) will reduce the number of levels.
\item With a larger D-Bucket,   the last mile search tail latency can be negatively affected.
\item A larger D-Bucket may also increase the search time of a range query since the key-value pairs are not sorted in a bucket.
\item However, as shown in Section \ref{sec-model-simplification}, a larger D-Bucket size can simplify data distribution, reduce prediction errors, and potentially improve lookup efficiency ($E_{\text{lookup}}$).
\end{itemize}

A higher initial fill ratio can reduce memory overhead, but it may only be suitable for read-most workloads. For insert-intensive workloads, maintaining the same D-Bucket size with a higher initial fill ratio results in fewer empty slots, potentially leading to longer search distances and more frequent D-Bucket splits.

\subsubsection{The hint design}
To accelerate the in-Bucket lookup process, BLI introduces \textit{Hint-assisted in-Bucket inserts (H-inserts)} and \textit{Hint-assisted in-Bucket lookups (H-lookups)} using \textit{hint functions}. A hint function suggests a preferred slot for the key to be looked up or inserted. One possible hit is to use a simple hash function. During an H-insert, if the hint-suggested slot is occupied, BLI sequentially checks from that slot until an empty one is found. Similarly, an H-lookup follows this sequential search if the suggested slot is occupied. Figure \ref{fig-architecture-base} shows an example hint function $h(k) = k \% C$, where $k$ is the input key and $C$ is the number of slots in a D-Bucket.

Both the "model" concept in learned indexes and our proposed hint functions map keys to their potential positions. The primary differences between them are as follows:
\begin{itemize}
\item Hint functions are not required to be monotonic since no enforced sorted order is needed.
\item Hint functions may not be derived from data distribution learning; instead, pre-defined hint functions like MOD hash, murmur hash, or CL hash can be used.
\item If the hinted slot is occupied by a non-target key-value pair, BLI continues searching forward, and any available empty slot can be used for insertion.
\end{itemize}

Currently, BLI employs a simple hash function as the hint because they result in short "searching forward" distances in our evaluation. However, a hint design with a bounded search distance may further improve BLI's efficiency.

\subsection{S-Buckets: adding unsortedness in Segments}

\begin{figure}[hbtp]
\centering
\includegraphics[width=0.3\textwidth]{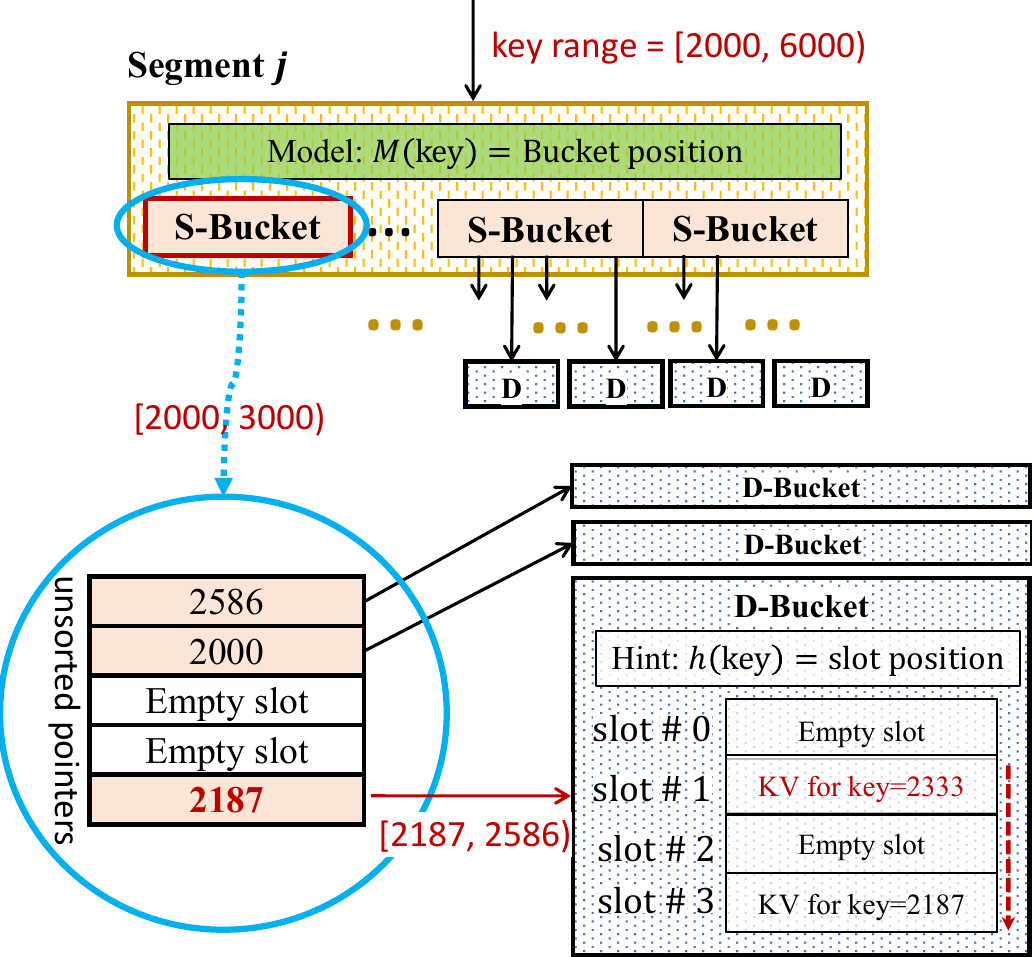}
\caption{\label{fig-sbuck} Structure of an S-Bucket}
\end{figure}

Organizing a sorted linear array of key-value pairs into D-Buckets introduces new opportunities and trade-offs. In BLI, each Segment also uses a sorted linear array to place child pointers. This raises the question of whether replacing these sorted child pointers with Buckets would also be advantageous. If so, what new design considerations arise?

We propose replacing the child pointer array with a list of \textit{Segment Buckets} or \textit{S-Buckets}, as shown in Figure \ref{fig-sbuck}. S-Buckets within a Segment are sorted by key range, while the child pointers within an S-Bucket remain unsorted. Empty slots within an S-Bucket allow for future key range updates. Similar to D-Buckets, we define the initial fill ratio as the fill ratio when generating a new S-Bucket. The number of S-Buckets for a Segment depends on the number of child pointers, the initial fill ratio, and the number of slots within an S-Bucket.  Consequently, the learned model should predict the position of an S-Bucket rather than a single child pointer. BLI can then identify the target S-Bucket near the predicted location and search for the expected child pointer within that S-Bucket.

\subsubsection{The hint design}
Similar to D-Buckets, introducing unsortedness in S-Buckets could improve insertion efficiency, while the lookup efficiency must be maintained in some way. However, a fundamental difference exists: a child pointer corresponds to a key range, whereas a key-value pair represents a single data point. If we want to support hints for S-Buckets, we should ensure that all keys within the key range of an S-Bucket are suggested to positions close to each other or "clustered." In other words, the hint function should be somewhat order-preserving. As a result, the aforementioned hash functions are no longer feasible.

We propose three possible designs for S-Buckets to improve lookup efficiency:
\begin{enumerate}
    \item \textit{Linear regression hint.} We can utilize a linear regression formula as a hint for each S-Bucket. Unlike learned regression models, when the expected slot is occupied during insertion, linear regression hints do not enforce order and instead place the child pointer in the next available slot.
    \item \textit{HOPE hint.} HOPE \cite{zhang2020order} is a High-speed Order-Preserving Encoder, a fast dictionary-based compressor that encodes arbitrary keys while preserving their order.
    \item \textit{S-Bucket size cap.} Rather than using hints, we can place the entry in the first available slot during insertion and cap the S-Bucket size to ensure fast lookup.
\end{enumerate}

Linear regression hints can become less "order-preserving" with more insertions, while HOPE introduces additional retraining overhead for new insertions. Designing hints for key ranges is more challenging than for single data points. In BLI, we adopt the third design because it is easy to implement and maintain.

Under the "S-Bucket size cap" choice, there are two possible implementations to represent a key range: 1) Record both the minimum and maximum key along with each child pointer; 2) Record the minimum key as the representative along with each child pointer. When looking up a key, the latter implementation requires visiting all representatives to determine the target (i.e., the largest representative smaller than or equal to the lookup key). In contrast, the former can stop early at the pointer whose minimum and maximum keys cover the lookup key. However, updating the maximum key of an S-Bucket could propagate to the Segment it belongs to and all parent Segments upwards. For ease of maintenance, BLI records only the minimum key.

\subsubsection{S-Bucket size and the initial fill ratio}

Currently, BLI caps the S-Bucket size to one or two cache lines. For instance, if each entry's minimum key is 8 bytes, each pointer is 8 bytes, and one cache line is 64 bytes, then the S-Bucket size would be capped at 8 or 16 entries. Similar to D-Buckets, larger S-Bucket sizes can result in greater model fanouts ($N$) and reduced prediction errors. However, larger S-Buckets could also degrade lookup performance. Regarding the fill ratio, its influence is similar to that on D-Buckets and should be determined by workload patterns.

\section{BLI Operations\label{sec-algorithm}}

In this section, we will discuss the major operations in BLI, including lookup, insert, bulk loading, and concurrency support.

\subsection{Lookup and range query\label{sec-bli_lookup}}

A series of root-to-leaf segment lookups are used to look up a key in BLI, followed by a D-Bucket exact-match lookup.\\

\subsubsection{Segment lookup: Bucket-based prediction}
The lookup operation in BLI starts at the root Segment. Within the Segment, the linear regression model is employed to predict the target S-Bucket rather than an individual slot. This linear model has errors. Thanks to segmentation-aware splitting, as we will discuss in the SMO section \ref{sec-seg-split}, the prediction errors are typically small. Moreover, as we have demonstrated in Section \ref{sec-resilience-model-inaccuracy}, because models predict on Bucket ID rather than the individual slot, the model simplicity is able to be tuned by S-Bucket size. With a simplified model, node fanouts can be further enlarged.\

Starting from the model-predicted S-Bucket, the process then searches among the neighboring S-Buckets to identify the one with the largest S-Bucket pivot that is smaller than or equal to the given lookup key. Upon identifying the target S-Bucket, an in-Bucket lookup is performed to find the entry with the largest pivot that remains smaller than or equal to the lookup key (i.e., the lower-bound key). Following the pointer of the target entry, the process proceeds to the next level. This procedure is iteratively performed at each level until the leaf D-Bucket is reached. 
\\

\subsubsection{D-Bucket lookup: H-Lookup \label{sec-H-lookup}}

D-Bucket lookups diverge from S-Bucket lookups, as they do not necessitate the ordering of entries. By capitalizing on this, a hint function can be utilized in D-Buckets, which serves as the starting search position. The choice of hint functions determines the entry distribution within a D-Bucket and thus affects the hint search distances (i.e., the number of slots visited starting from the hint slot). We evaluated different hint choices, including linear model, MOD hash, and CL hash hints.

\textbf{Endpoint Linear Models} To start with, the hint function can be defined as a linear model:
$$ h(k) = a k + b$$
In this formula, $a$ is the slope, and $b$ is the intercept. As we use hints to accelerate in-Bucket search, the overhead of computing the hint must be small. The endpoint linear model is computed in $O(1)$ time, using the pivot of both the target D-Bucket and its right neighboring D-Bucket.\

\textbf{Hash hints} Using endpoint linear models almost introduces no additional computing overhead, but its efficiency highly depends on the key distribution within the D-Bucket. The linear model can have more conflicts when the local key distribution is less linear (e.g., it has a smaller Pearson correlation coefficient).\

Therefore, we use hash functions to randomize the original key distribution and reduce the number of conflicts. An ideal hash function has the following properties: low latency, even distribution, and low collision rate \cite{yamaguchi2013hardware}. This paper evaluated three different hash functions, including MOD and CL hash. Any other hash functions or hint algorithms can be smoothly applied in the D-Bucket without modifying other components in BLI.

The simplest hash function we evaluated is MOD hash, $h(k) = k \% C$, where $C$ is the D-Bucket size. It has low computing overhead. On the other hand, if the keys have similar reminders after being divided by $C$, there will be many conflicts within the D-Bucket. If the key distribution is linear enough (i.e., has a small linear regression error), it can have a high chance of having this "similar reminder" data distribution. In fact, we observed many conflicts in one of the simple datasets called \texttt{covid}. Besides, CL hash is designed to run extremely fast on modern processors equipped with the Carry-less Multiplication (CLMUL) instruction set.

\begin{figure}[htbp]
\centering
\includegraphics[width=0.45\textwidth]{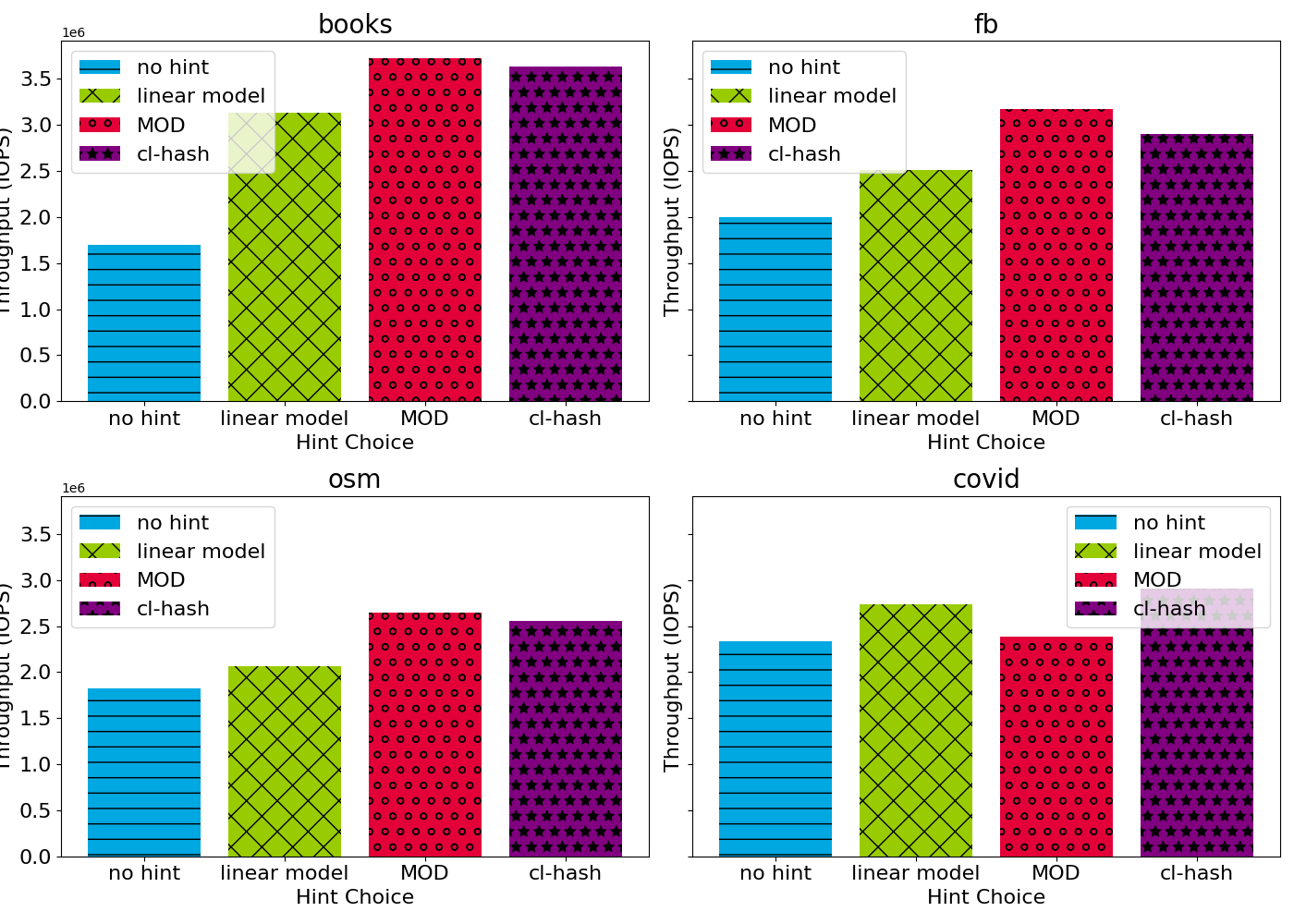}
\caption{\label{fig-hint-throughput} Throughput under different hint choices with the read-only workload.}
\end{figure}

We evaluated two hashing functions and found that overall, CL hash was able to randomize the data distributions, as shown in Figure \ref{fig-hint-throughput}. Note that MOD hash has almost no improvement when compared with the no hint case on \texttt{covid}, which is an easy and linear dataset.

\subsubsection{Range query} For range queries, BLI starts by identifying the target D-Bucket of the start key, and continues visiting the subsequent D-Buckets untils the number of visited key-value pairs reaches the required amount. After that, BLI use multiple thread workers to copy the key-value pairs in each D-Bucket to a temporate buffer, and sort these key-value pairs in parallel. 

Given a query of $n$ key-value pairs, the number of D-Buckets involved $N$ is $O(\frac{n}{wC})=O(n)$, where $w$ represent the number of thread workers, and  $C$ represents the D-Bucket size. For each D-Bucket, the sorting takes $O(C \log C) = O(1)$ time. Therefore, the amortized in-Bucket query cost is $O(\frac{N \cdot 1 + n}{n})=O(1)$. On the other hand, D-Bucket size can affect the tail latency as it takes a longer time to create or sort a larger D-Bucket. 

\subsection{Bottom-up insert and simple SMOs\label{sec-insert-and-smo}}
Within BLI, insertions occur in a bottom-up fashion, and they may trigger SMOs, including D-Bucket split, S-Bucket split, Segment split, Segment merging, Segment scaling, and model retraining. We will first discuss the insertion into a non-full D-Bucket, which doesn't necessitate any SMOs. Following that, we will introduce the D-Bucket split first and then the rest of the SMOs. It is worth noting that the rest of the SMOs in BLI are combined in one operation and executed in linear time with respect to the number of entries in the S-Bucket, which is highly efficient.

\subsubsection{Insert in non-full D-Bucket: H-insert}
In order to insert a key, a lookup is initially performed to identify the target D-Bucket whose key range contains this key. If the target D-Bucket is not filled to its capacity, the new key-value pair is inserted into this D-Bucket. Similar to a lookup, BLI first checks whether the hinted slot is already occupied. If not, it continues forward to locate the first empty slot.

The unsorted nature of D-Buckets is highly efficient because it eliminates the need for strict-order-preserving operations (e.g., shifting in ALEX and chaining in LIPP). First, no shifting is required since any empty slots can be used for insertion. Second, BLI does not need chaining because the Buckets can act as a buffer to delay conflicts. During insertion in LIPP, the incoming key-value pair is inserted into the model-predicted slot if it is empty. However, if the single predicted slot is already occupied, LIPP performs chaining to accommodate the new key-value pair in a new node. In contrast, BLI does not stick to a specific slot for insertion. Only after all empty slots within a D-Bucket are occupied does BLI begin to perform a "conflict-resolving" operation, which we call the D-Bucket split.

\subsubsection{D-Bucket split \label{sec-dbuck-split}}
When the D-Bucket targeted for insertion is full, it is evenly split into two distinct D-Buckets, with each new D-Bucket accommodating half of the key-value pairs. This process is executed in linear time with respect to the number of entries in the D-Bucket. We employ Hoare's quick selection algorithm\cite{hoare1961algorithm} to identify the median key. Subsequently, the key-value pairs with keys less than or equal to the median key are placed in the first newly created D-Bucket, while the rest of the key-value pairs are allocated to the second D-Bucket. In addition, the new key-value pair will be inserted into one of the two new D-Buckets. \

\subsection{Combined SMOs in linear Time\label{sec-seg-split}}
After the D-Bucket split, the (\texttt{key}, \texttt{pointer}) pair representing the newly generated D-Buckets needs to be inserted into their parent Segment. This insertion can trigger additional SMOs, including Segment scaling, Segment split, Segment merging, and model retraining. \textit{Segment scaling} involves enlarging the Segment capacity while maintaining the current model. \textit{Segment split} divides a Segment into multiple ones when the key distribution within the Segment is no longer linear enough, necessitating re-segmentation. \textit{Segment merging} combines neighboring segments that fit well into a single linear model. \textit{Model retraining} relearns the distribution of entries in the current Segment.

Initially, the "current Segment" refers to the parent Segment of the old D-Buckets being split. Additional SMOs may also propagate upwards, affecting the parent of the current Segment. In the rest part of this section, we will focus on a single SMO occurring within the current Segment.

The trigger conditions for these SMOs must be carefully designed. We classify them into passive SMOs and active SMOs. Passive SMOs are triggered due to capacity limits when a D-Bucket, S-Bucket, or Segment is full. Conversely, active SMOs may be triggered due to performance degradation. For example, if the current Segment is not linear enough and has large prediction errors, model retraining, Segment split, or Segment scaling can be performed to address this issue. If the overall distribution of neighboring D-Buckets is linear, Segment merging may be initiated. The trigger conditions determine the SMO triggering frequency. If SMOs are triggered too frequently, it interferes with the normal lookup and insert operations; if SMOs are triggered too rarely, the Segment layout may be suboptimal. BLI triggers one big SMO when an S-Bucket is full, which makes the SMO trigger less frequently than when a D-Bucket is full and more frequently than when a Segment is full. Notably, the one big SMO efficiently combines Segment scaling, Segment split, model retraining, and neighbor Segment merging together.

\begin{algorithm}[htbp]
    \caption{Combined SMO for node scaling, model retraining, node split, and node merge\label{algorithm:com-smo}}
    \SetKwInOut{Input}{Input}
    \SetKwInOut{Output}{Output}
    \Input{Current segment $B$, new entries $E_{\text{in}}$}
    \Output{Entries of newly generated segments $E_{\text{out}}$}
    \BlankLine
    
    \If{average number of SMOs for current bucket among neighboring segments $< \theta$}{
        ReSegment($B$, $E_{\text{in}}$)\;
    }
    \Else{
        MergeNeighbors($B$, $E_{\text{in}}$)\;
    }
\end{algorithm}

\begin{algorithm}[htbp]
    \caption{ReSegment Function}
    \SetKwFunction{ReSegment}{ReSegment}
    \SetKwFunction{MergeEntries}{MergeEntries}
    \SetKwFunction{Segmentation}{Segmentation}
    
    \ReSegment{$B$, $E_{\text{in}}$}{
        keys $\leftarrow$ MergeEntries($B$.entries, $E_{\text{in}}$)\;
        $(C_1, C_2, \ldots, C_n) \leftarrow$ Segmentation(keys)\;
        
        \For{$i \leftarrow 1$ \KwTo $n$}{
            segment$_i \leftarrow C_i$.model + list of $\left\lceil\frac{|C_i.\text{keys}|}{\text{fillRatio}}\right\rceil$ empty S-Buckets\;
            \ForEach{entry $x \in C_i$}{
                segment$_i$.ModelInsert($x$)\;
            }
        }
    }
\end{algorithm}

\begin{algorithm}[htbp]
    \caption{MergeNeighbors Function}
    \SetKwFunction{MergeNeighbors}{MergeNeighbors}
    
    \MergeNeighbors{$B$, $E_{\text{in}}$}{
        leftBoundary $\leftarrow B$\;
        \While{leftBoundary is bounded by GreedyCorridor}{
            leftBoundary $\leftarrow$ left neighboring segment of leftBoundary\;
        }
        rightBoundary $\leftarrow B$\;
        \While{rightBoundary is bounded by GreedyCorridor}{
            rightBoundary $\leftarrow$ right neighboring segment of rightBoundary\;
        }
        oldLCA $\leftarrow$ least common ancestor of leftBoundary and rightBoundary\;
        pivots $\leftarrow$ all D-Bucket pivots of the LCA sub-tree $\cup$ $E_{\text{in}}$\;
        newLCA $\leftarrow$ bottomUpPropagate(pivots)\;
        oldLCA $\leftarrow$ newLCA\;
    }
\end{algorithm}






     
    

\SetKwProg{Fn}{Function}{:}{end}
\LinesNumbered
\Fn{\MergeNeighbors{$B$}}{
    $leftBoundary \leftarrow B$;\\
    \While{$leftBoundary$ is bounded by $GreedyCorridor$}{
        $leftBoundary \leftarrow$ the left neighboring Segment of $leftBoundary$;
    }
    $rightBoundary \leftarrow B$;\\
    \While{$rightBoundary$ is bounded by $GreedyCorridor$}{
        $rightBoundary \leftarrow$ the right neighboring Segment of $rightBoundary$;
    }
    $oldLCA \leftarrow$ the least common ancestor of $leatBoundary$ and $rightBoundary$;\\
    $pivots \leftarrow$ all D-Bucket pivots of the LCA sub-tree and also $E_{\text{in}}$;\\
    $newLCA \leftarrow$ bottomUpPropagate(keys);\\ 
    $oldLCA \leftarrow newLCA$;\\
}

We categorize the aforementioned SMOs into two groups: Segment scaling, model retraining, and Segment split are bundled in the function \ReSegment{}, as they may be triggered when one Segment is not linear enough; \MergeNeighbors{} constitutes the second category, focusing on simplifying the node layout. As shown in Line 1 of Algorithm \ref{algorithm:com-smo}, we use an indicator to decide whether to trigger \ReSegment{} or \MergeNeighbors{}. This indicator is based on the average number of SMOs for the current Segment to be processed among its neighboring Segments. If the average number of SMOs is higher than a threshold, we proactively trigger node merging. This indicator can be easily replaced by other factors, and we can set different indicators to control the frequency of node merging.

\subsubsection{Node scaling, model retraining, and node split}

\textsc{ReSegment} takes the current segment $B$ and new entries $E_{\text{in}}$ as input. The new entries are key-pointer pairs for the two new D-Buckets, and the current segment is the parent segment of these D-Buckets. First, it merges the pivots of $E_{\text{in}}$ and the entries from all S-Buckets into a single sorted array \texttt{keys} using merge-sort in linear time. Then, a single-pass greedy segmentation algorithm \textsc{GreedyCorridor}~\cite{neumann2008smooth} is performed on \texttt{keys}. Based on the segmentation result, the current segment can either split into multiple segments or remain as a single segment. Finally, it generates new segments according to the predefined fill ratio (e.g., 60\% full), during which node scaling and model retraining occur.

The outputs of a \ReSegment{} are the new entries, each representing a newly generated Segment. The number of new entries or Segments is greater than or equal to one. These new entries will be propagated to the upper level to be inserted. The insertion process can also trigger SMOs if there are not enough empty slots at the parent Segment level. The propagation continues until it can be inserted at a level or until it reaches the root. If there is still more than one Segment at the root level, a new level will be generated and set as the new root.

\subsubsection{Node merging}
In \MergeNeighbors{}, BLI needs to identify which D-Buckets are to be merged in a lightweight manner. Our proposed identification algorithm grows the Segment incrementally, and it works as follows: at the leaf-node level, we start at the newly split D-Bucket and expand in both left and right directions to check whether the neighbor D-Buckets can be merging candidates. We terminate if we reach a D-Bucket that is not supposed to be the same Segment. We reuse the Greedycorridor segmentation algorithm, with the same predefined error bound as in bulk loading and node split phase, to determine if the neighbor D-Bucket can fit into the same Segment. There are three major reasons to choose Greedycorridor as the node merging algorithm: 1) the time complexity of Greedycorridor is linear regarding the number of neighbor D-Bucket visited, which is lightweight; 2) Greedycorridor works by sequentially transverse each key and stops when the slope change is too much. During the linear transversal, we can incrementally update the linear model; 3) It makes the node splitting and merging algorithms consistent with each other, with the same hyper-parameter required (i.e., the error bound), which makes it easier to tune BLI performance. After identifying the left and right boundary D-Buckets, BLI finds their least common ancestor $oldLCA$. After re-building the sub-tree whose root is $oldLCA$ in a bottom-up manner, $oldLCA$ is finally replaced by $newLCA$. Similar to bulk loading, this bottom-up propagation traverses the D-Buckets in a single pass, which is highly efficient.

\subsection{Bottom-up single-pass bulk loading}

BLI also supports an efficient bottom-up bulk-loading algorithm. Given a sorted array of key-value pairs, BLI partitions all key-value pairs evenly into Buckets in a single pass. Based on the initial fill ratio and Bucket size, we compute the number of key-value pairs in each Bucket as Bucket size times the initial fill ratio. After that, BLI put adjacent key-value pairs into Buckets accordingly. However, if we vary Bucket size or the initial fill ratio among different Buckets, it is possible to achieve a more linear key distribution.  Similar to the existing bottom-up approaches, the subset size at each upper level diminishes exponentially. After bulk loading, the BLI structure is a balanced tree without long and skewed paths.

\subsection{Lock-free concurrency support\label{sec-concurrency}}
\noindent We support lock-free single-producer-multi-consumer (SPMC) concurrency, where one dedicated thread can insert key-value pairs into the index while not blocking lookup from other threads. Meta claimed that the read-to-write ratio in their KV cache is 30:1\cite{atikoglu2012workload}, which matches well with our design, which has one writer and several readers. Moreover, the SPMC scenarios are especially useful for sharded key-value stores\cite{didona2019size}. Sharding is the mechanism that splits the key range into multiple partitions (or shards). In each shard, to employ BLI under the SPMC scenario, we can have one dedicated writer to handle the insert process while all readers from other sessions can run in parallel without blocking. 

BLI supports SPMC in a lock-free manner. This can be achieved by 1) using a valid bit for each entry with a specifically designed update order and 2) using a Read-Copy-Update (RCU) for all SMOs. More specifically, to ensure data consistency, updating the entry's valid bit happens after the data insertions. When inserting an entry into a non-full node, BLI first inserts it at an empty slot, then sets its valid bit to 1. If the inserted entry is the smallest one of the Bucket, BLI will update the Bucket's pivot as the last step. For SMOs, including the D-Bucket split and the combined SMO, we use RCU to avoid locking.

\begin{figure}[htbp]
\centering
\includegraphics[width=0.48\textwidth]{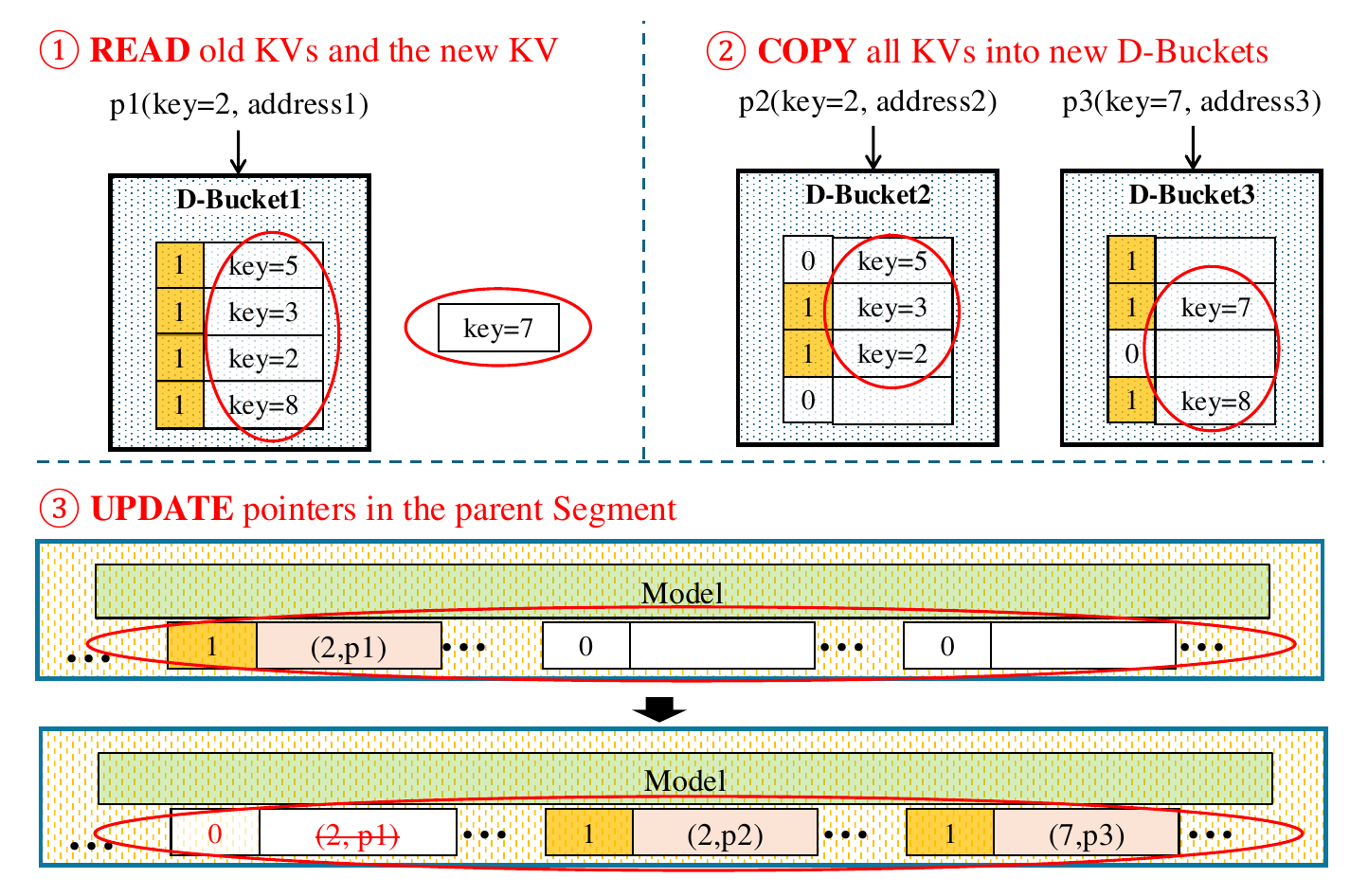}
\caption{\label{fig-D-Bucket-RCU} The RCU process for D-Bucket splits in BLI. }
\end{figure}

Figure \ref{fig-D-Bucket-RCU} illustrates the RCU process for D-Bucket splits. In step 1, BLI \textit{reads} all key-value pairs from the old D-Bucket (D-Bucket1) to be split, which includes the key-value pairs with keys 5, 3, 2, and 8, along with the new key-value pair to be inserted (key = 7). In step 2, BLI calculates the median key from this set of key-value pairs, which is 5. All key-value pairs (including the new key-value pair to be inserted) with keys less than or equal to 5 are \textit{copied} into the first new D-Bucket (D-Bucket2), while the remaining key-value pairs are \textit{copied} into the second new D-Bucket (D-Bucket3). In step 3, the key-pointer pairs for the two new D-Buckets (i.e., pair p2 for key = 2, address = address1, and pair p3 for key = 5, address = address3) are inserted into the parent Segment. In this example, since there are empty slots available to store the new key-pointer pairs, no further SMOs are triggered. However, if the new key-pointer pairs cause an SMO due to capacity limits in the Segment, BLI will invalidate the key-pointer pair to the old Bucket only \textit{after} all SMOs and propagation (e.g., ReSegment() or MergeNeighbors() for the affected Segments at different levels) are complete. Other SMOs follow similar RCU procedures with the process for D-Bucket splits.

\section{Evaluation\label{sec-evalution}}

\noindent In this section, we first compared the performance of BLI with several state-of-art in-memory learned indexes, including ALEX\cite{ding2020alex}, LIPP\cite{wu2021updatable}, PGM\cite{ferragina2020pgm}, and FINDex\cite{li2021finedex}. ALEX and LIPP are designed for a single-core machine and did not initially work in a multi-thread environment. Thus, we used the implementation from \cite{wongkham2022updatable}. The concurrent versions are denoted as ALEX+ and LIPP+. After that, we adopted drill-down analyses on different factors to validate the effectiveness and efficiency of our design components.

\subsection{Evaluation set up}
\noindent We implemented BLI in C++ and performed evaluations on an Ubuntu Linux machine equipped with an Intel(R) Xeon(R) CPU E5-2670 v3 @ 2.30GHz and 128GB RAM.  We used 8B keys and 8B payloads in all our experiments. To have a comprehensive evaluation, we compare the following metrics: 1) The overall lookup/insert performance, 2) The time breakdown of lookup/insert operations, and 3) The memory consumption.\\

The dataset we used followed the GRE benchmarking paper\cite{wongkham2022updatable}. We chose \texttt{books}, \texttt{fb}, and \texttt{osm} datasets as representatives because they represent three different "hardness" (a measurement of how linear they are) patterns defined in GRE. They are identified as easy, medium, and hard datasets, respectively. Regarding the workload, to investigate different scenarios comprehensively, we studied various read-write ratios from 0:1, 1:9, 2:8, ... to 1:0, which covers read-only, insert-only, read-write ($\text{read}:\text{write} = 1:1$), and read-write cases. Given that each dataset we used consists of 200M keys in total, we randomly bulk-loaded 100M key-value pairs and performed the rest of the 100M read or insert operations according to the read-insert ratio.

\begin{figure}[htbp]
\centering
\includegraphics[width=0.44\textwidth]{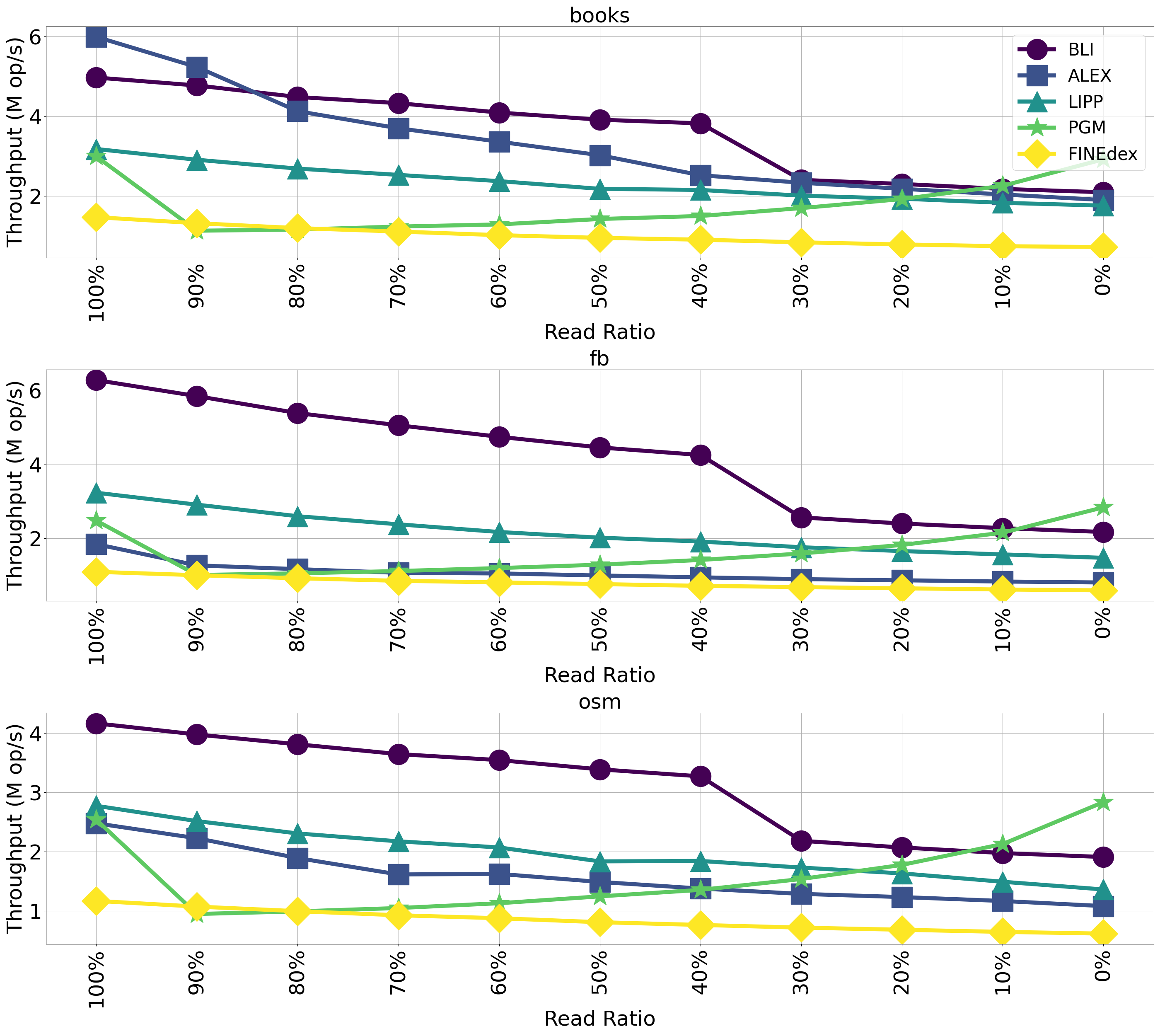}
\caption{\label{exp:overall} Throughput under different read-write ratios.  }
\end{figure}

\subsection{Overall throughput comparison}
Figure \ref{exp:overall} compares the throughputs of SOTA learned indexes and BLI across various read-write ratios. BLI outperformed other learned indexes in most workloads and datasets. As shown in the figure, PGM achieves the highest throughput in insert-only workloads because it accommodates insertions similarly to LSM-tree compaction, and the primary advantage does not stem from the learned index design itself \cite{wu2021updatable}. Notably, the figure illustrates a sharp decline in PGM's performance with even a small read ratio (10\% or 20\%). Consequently, PGM is excluded from further analysis.

In read-only workloads, BLI showed improvements of -30.9\%, +37.86\%, and +15.35\% compared to the best-performing SOTA index on each respective dataset. BLI consistently outperformed all SOTA indexes on medium and hard datasets. Conversely, BLI underperforms relative to ALEX in read-only/read-heavy workloads on the \texttt{books} dataset due to its extremely linear key distribution. For such "easy" datasets, ALEX's range partitioning, guided by its cost estimation model, can terminate early in linear distributions, resulting in fewer levels. 

In insert-only workloads, BLI demonstrated improvements of 10.27\%, 47.79\%, and 40.35\% compared to the \textit{best} SOTA index on \textit{each} dataset (i.e., ALEX, LIPP, and LIPP, respectively). In a more realistic workload scenario with a read-write ratio of $1:1$, BLI surpassed the best SOTA learned indexes by 29.43\%, 121.60\%, and 84.43\% on the respective datasets (i.e., ALEX, LIPP, and LIPP, respectively). These results affirm BLI's consistent performance advantage over other SOTA learned indexes as an updatable learned index, especially under realistic workloads.

\subsection{S-Bucket and D-Bucket sizes}
\begin{figure*}[htbp]
    \centering
    \includegraphics[width=\textwidth]{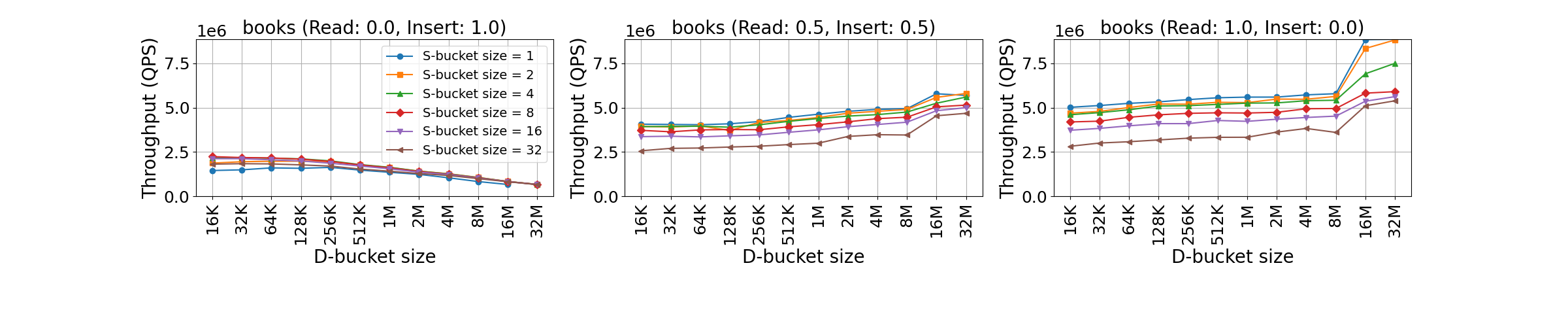}
    \caption{Throughput with different S-Bucket and D-Bucket size on \texttt{books}}
    \label{exp:buck-size-throughput}
\end{figure*}

We evaluated how S-Bucket and D-Bucket sizes affect the throughput under different workloads. Figure \ref{exp:buck-size-throughput} illustrates that different workloads prefer different S-Bucket and D-Bucket sizes on \texttt{books}, with similar trends observed across the other two datasets.

For insert-only workloads, smaller D-Buckets are preferred because 1) insertions can trigger SMOs, including D-Bucket splits, and smaller D-Buckets reduce the time needed to generate a new D-Bucket, and 2) the maximum hint search distance is bounded by D-Bucket size. As a result, smaller D-Buckets can lead to better $E_{insert}$ and $E_{lookup}$. Regarding S-Bucket sizes under insert-only workloads, BLI's throughput achieves its peak value when the S-Bucket size is set to 4; either increasing or decreasing the S-Bucket size reduces BLI's throughput. This is because when the S-Bucket is too small, SMOs will be triggered more frequently, as SMOs occur when an S-Bucket becomes full, which can impact Model Fanout $N$ and Insert Efficiency $E_{\text{insert}}$. Conversely, when the S-Bucket is too large, finding the target entry within the S-Bucket takes longer since entries within an S-Bucket are unsorted, which negatively affects $E_{\text{lookup}}$ within the S-Bucket.

For read-only workloads, BLI prefers larger D-Bucket sizes and smaller S-Bucket sizes. With no new key-value pairs being inserted, there is no overhead associated with generating new D-Buckets. Additionally, given the same number of key-value pairs, larger D-Bucket sizes lead to fewer D-Buckets, simplifying the tree structure and increasing Model Fanout $N$. The sudden performance boost from a D-Bucket size of 16M to a D-Bucket size of 32M is due to a reduction in the number of levels by one. For S-Buckets, the absence of SMO overhead allows smaller S-Buckets to reduce $E_{\text{lookup}}$ within each S-Bucket while not affecting $E_{\text{insert}}$, as previously discussed.

\begin{figure}[htbp]
    \centering
    \includegraphics[width=0.5\textwidth]{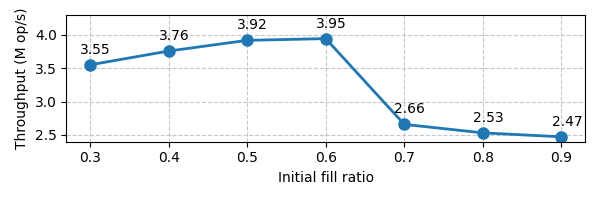}
    \caption{Throughput with different initial fill ratios on \texttt{books}}
    \label{exp:fill-ratio}
\end{figure}

\subsection{Initial fill ratio}
From our experimental observations, D-Buckets predominantly determine the memory size. In other words, the memory overhead of Segments is neglectable (i.e., less than 1\%). The memory consumption of all D-Buckets is determined by the initial fill ratio, which is the fill ratio when allocating new Buckets. With a smaller initial fill ratio, the memory consumption for Buckets is larger, which is bad for the Memory Overhead $O\text{mem}$. In terms of performance, BLI performs best when the initial fill ratio is 60\%. With a smaller fill ratio, more D-Bucekts are needed to place the same set of key-value pairs, which can harm Model Fanout $N$; with a larger fill ratio, the hint error distance can be bad, which can harm $E_{\text{lookup}}$.

Our experimental observations reveal that D-Buckets predominantly determine memory size, while the memory overhead of Segments is negligible (i.e., less than 1\%). The memory consumption of all D-Buckets is determined by the initial fill ratio, which represents the fill ratio at the time of allocating new Buckets. Figure \ref{exp:fill-ratio} demonstrates BLI's performance under various initial fill ratios. A lower initial fill ratio increases memory consumption, adversely impacting Memory Overhead $O_{\text{mem}}$.  In terms of performance, BLI achieves optimal results with an initial fill ratio of 60\%. A lower fill ratio requires more D-Buckets to store the same set of key-value pairs, which can negatively impact Model Fanout $N$. Conversely, a higher fill ratio can lead to greater hint error distances, detracting from Lookup Efficiency $E_{\text{lookup}}$.

\subsection{Range query}
To evaluate the performance of range queries, we bulk-loaded each index using a dataset of 200M keys and ran a read-only scan workload. A random start key was chosen, and the number of keys scanned was varied from 10 to $10^6$. The results show that BLI's best throughput reached 11.92 million keys/s, while ALEX achieved a significantly higher throughput of 33.22 million keys/s. BLI’s lower performance in range queries is due to the need to sort key-value pairs within each D-Bucket before scanning. With a smaller D-Bucket size, the sorting overhead decreases. However, BLI sacrifices range query performance to 1) improve point query performance and 2) facilitate lock-free concurrency.

It is important to note that these evaluations are conducted in memory. If the leaf data were stored on disk, the sorting overhead would become negligible, as I/O would then be the primary bottleneck. Future work includes optimizing BLI for scenarios where data exceeds main memory capacity.

\subsection{Time breakdown}
We present the time breakdown of BLI execution in Table \ref{table-time-breakdown}, based on experiments conducted on the \texttt{fb} dataset with a read-write ratio of $1:1$. In this workload, reads consume 40.19\% of the execution time, while insertions account for 59.90\%. Within the read operations, only 13.49\% of the time is spent traversing to the target D-Bucket. For insertions, the overhead of allocating new data comprises just 25.27\% of the insertion time.

\begin{table}[]

\caption{BLI time breakdown on \texttt{fb}}
\begin{tabular}{|c|l|}
\hline
\label{table-time-breakdown}
\multirow{2}{*}{\centering get() 40.91\%} & Segment lookup 13.49\% \\ \cline{2-2} 
                                          & D-Bucket lookup 86.51\% \\ \hline
\multirow{2}{*}{\centering put() 59.90\%} & Insert 74.73\%          \\ \cline{2-2} 
                                          & Memory management 25.27\%             \\ \hline
\end{tabular}
\end{table}



















\subsection{Concurrency}
\begin{figure}[H]
\centering
\includegraphics[width=0.45\textwidth]{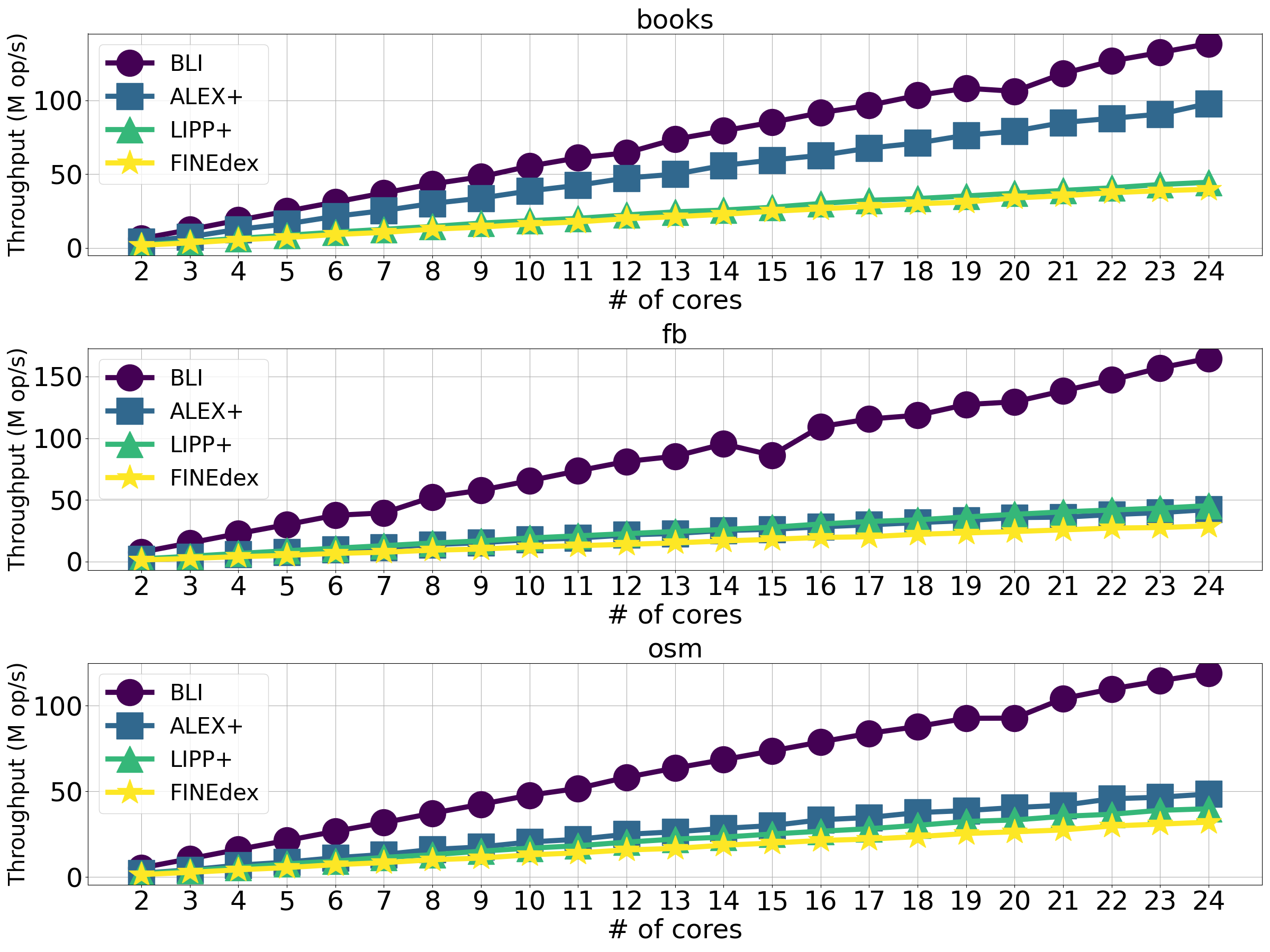}
\caption{\label{exp:mrsw} Throughput with different number of cores}

\end{figure}

To compare the performance of BLI and other indexes under read-intensive MRSW workloads, we set the read-write ratio to 7:3. We vary the number of threads (or cores) from 2 to 24. For $n$ cores, we evenly distribute all read operations to $n-1$ reader threads. Figure \ref{exp:mrsw} illustrates the results. Thanks to the lock-free concurrency, BLI outperforms all other SOTA learned indexes by up to 3.91x in Figure \ref{exp:mrsw}.


\subsection{Bulk load}  
Table \ref{exp-bulk-load} compares BLI's bulk loading throughput with ALEX and LIPP on the \texttt{books} dataset, with similar results observed on the other two datasets. Due to its bottom-up mechanism, BLI achieves significantly higher throughput, outperforming ALEX and LIPP by 557.19\% and 149.34\%, respectively.

\begin{table}[]
\caption{Bulk load throughput on \texttt{books}}
\begin{tabular}{|l|l|l|}
\hline
\label{exp-bulk-load}
\textbf{dataset} & \textbf{index} & \textbf{throughput (M op/s)} \\ \hline
books            & BLI            & 16.18                        \\ \hline
books            & ALEX           & 2.53                         \\ \hline
books            & LIPP           & 6.77                         \\ \hline
\end{tabular}
\end{table}


\section{Conclusion\label{sec-conclusion}}
We presented BLI, an in-memory index that employs a "globally sorted, locally hinted" architecture. This design incorporates hint-assisted Buckets and supports lock-free concurrency. BLI adopts incremental learning, allowing for efficient updates to underlying linear models. Performance evaluations confirm that BLI outperforms state-of-the-art learned indexes with up to a 2.21x throughput speedup. These results open avenues for future research in optimizing learned index structures.



\bibliographystyle{ACM-Reference-Format}
\bibliography{main}

\end{document}